\documentclass[aps,prb,amsmath,amssymb,amsfonts,twocolumn,nofootinbib]{revtex4}
\usepackage{graphicx}
\usepackage{dcolumn}
\usepackage{bm}
\usepackage{amsmath}
\usepackage{amssymb}
\usepackage{color}

\usepackage{soul}

\newcommand{\be}{\begin{equation}}
\newcommand{\ee}{\end{equation}}

\newcommand{\bea}{\begin{eqnarray}}
\newcommand{\eea}{\end{eqnarray}}

\begin{document}

\title{Spatiotemporal buildup of the Kondo screening cloud}

\author{M. Medvedyeva}
\affiliation{Department of Physics, Georg-August-Universitaet Goettingen,
Friedrich-Hund-Platz 1, 37077 Goettingen, Germany}
\author{A. Hoffmann}
\affiliation{Department of Physics, Arnold Sommerfeld Center for Theoretical Physics,
Ludwig-Maximilians-Universitaet, Theresienstrasse 37, 80333 Muenchen, Germany}
\author{S. Kehrein}
\affiliation{Department of Physics, Georg-August-Universitaet Goettingen,
Friedrich-Hund-Platz 1, 37077 Goettingen, Germany~}

\begin{abstract}
We investigate how the Kondo screening cloud builds up as a function of space and time.
Starting from an impurity spin decoupled from the conduction band, the Kondo coupling is
switched on at time $t=0$. We work at the Toulouse
point where one can obtain exact analytical results for the ensuing spin dynamics at both zero
and nonzero temperature~$T$. For $t>0$ the Kondo screening cloud starts building up in the wake of
the impurity spin being transported to infinity. 
In this buildup process the impurity spin--conduction band spin susceptibility
shows a sharp light cone due to causality, while the corresponding correlation function 
has a tail outside the light cone. At $T=0$ this tail has a power law decay as a function
of distance from the impurity, which we interpret as due to initial entanglement in the Fermi sea.   
 \end{abstract}

\maketitle

\section{Introduction}

The world around us is essentially a non-equilibrium system. There is still a lot to understand on how excited systems evolve with time. For example, how do information and correlations spread in a non-equilibrium system? Perhaps the easiest setup to consider is a quantum quench~\cite{quench}: One prepares a system in the ground state of some Hamiltonian, and then suddenly changes the Hamiltonian so that the initial state is no longer an eigenstate. Therefore the time evolution of the system becomes non-trivial. One aspect of this time evolution is that initially unentangled parts of the system can become entangled. 

From the semiclassical point of view this entanglement propagates via quasiparticles.~\cite{Cal07} 
For example, a perturbation acting at one point of the system leads to excitation of quasiparticles. These propagate to neighboring regions, perturb the system locally, and excite new quasiparticles which carry the information about the initial perturbation further and further. If we assume a finite speed of the quasiparticles, the propagation of the information can be described by an effective light cone -- the information about the excitation has reached the points inside the light cone, but not outside.

Historically, effective light cones in the dynamics of non-relativistic quantum many-body systems were first investigated in the context of quantum spin chains by Lieb and Robinson.\cite{LiebRobinson} They proved rigorously 
that certain commutators have a structure akin to relativistic field theory in the sense that they decay exponentially outside the light cone. Due to the importance of understanding the spread of
entanglement in quantum information processing and efficient numerical simulation methods, a lot of theoretical work has since then been done to generalize the original work by Lieb and Robinson
to other situations and more general questions.\cite{Nac,Has10,Bra06,Schuch} Numerically, the light cone effect has been seen in a number of lattice models.~\cite{numLC,Kollath2008} Recently, it was also observed
experimentally after a quench in a cold atomic gas with very good agreement with theoretical results.\cite{exp2}

In our paper we present the analytical study of the light cone effect in an exactly solvable model, namely the Kondo model at the Toulouse point. Specifically,
we consider a Kondo impurity coupled to the conduction band electrons at time $t=0$. It is well known that
the impurity spin degree of freedom is screened in equilibrium  by a Kondo screening cloud of 
conduction band electrons, which leads to a Kondo singlet ground state.\cite{Hew97} This Kondo screening has 
been the subject of intensive research over
many years, both experimentally and theoretically.\cite{Affleck} In our non-equilibrium setup we are interested in 
how this Kondo screening builds up as a function of space and time starting from an unentangled state at~$t=0$
\be
|\psi_{NEQ}\rangle = |\uparrow\rangle \otimes |{\rm FS}\rangle \ ,
\label{initialstate}
\ee
where $|{\rm FS}\rangle$ is the non-interacting Fermi gas. Among other results, we will see how the initial impurity
spin is transported to infinity in order to asymptotically obtain a Kondo singlet ground state.

Technically, the calculation proceeds by an exact mapping of the Kondo
model at the Toulouse point
to a quadratic Hamiltonian using bosonization and refermionization.~\cite{RNM,Guinea,Kehr1}
The properties of the Kondo model at the Toulouse limit represent well the general properties at the strongly coupled fixed point,~\cite{Aff95} even though the Toulouse limit corresponds to an anisotropic Kondo model. The generic behavior was shown in Ref.~\onlinecite{Hof01} -- the equilibrium spin correlations functions in the isotropic and anisotropic Kondo models are quite similar depending on the anisotropy parameter.
The quadratic model at the Toulouse point is simply a 
resonant level model, which is effectively relativistic with the Fermi velocity corresponding to the speed of light since
the conduction band Hamiltonian is linearized around the Fermi energy yielding a linear dispersion of the electrons in the conduction band. We derive the time-dependence of the creation and annihilation operators of the electrons and the impurity spin. This gives us a straightforward way to calculate the time and spatial dependence of the correlation functions for our quench setup. We will see that the commutator of two spins is zero outside the effective light cone, as it should be from causality considerations. On the other hand, the equal time correlations corresponding to the anticommutator exhibit a light cone with a non-zero tail outside the light cone, which decays
with a power law at zero temperature. 
Similar calculations were done in Ref.~\onlinecite{Guinea}, but without time-dependence. Related observations about the propagation of  excitations were obtained in Ref.~\onlinecite{SCqubits} in the context of information spread in a system of two qubits coupled via a conduction line. 

We conclude with a discussion how our work is connected to Lieb-Robinson bounds and discuss possible future directions of work.

\section{Model and formalism}

The presence of a localized spin-1/2 degree of freedom coupled to the conduction band gives rise to the Kondo effect -- the formation of
the Kondo screening cloud
around the unpaired spin, which screens the impurity spin.~\cite{Hew97}

{This behavior can be derived from the Anderson impurity model under the assumption that an unoccupied or doubly occupied impurity orbital is not energetically favorable. This assumption leads to the three-dimensional Kondo Hamiltonian
\be H = \sum_{k,\sigma} \epsilon(\vec{k})\,f^\dagger_{\vec{k}\sigma}f_{\vec{k}\sigma}+\sum_{i=x,y,z} J_i S_i s_i^{el}(\vec{r}=0),\ee
where $f^\dagger_{\vec{k}\sigma}$ and $f_{\vec{k}\sigma}$ are creation and annihilation operators of the conduction band electrons. $\vec{s}^{el}(\vec r=0)$ is the spin of the conduction band electron localized at the origin, $\vec{S}$ is the quantum impurity spin and $\vec{J}$ the coupling between the impurity and conduction band electron spins. It can be anisotropic: $\vec{J}=(J_{\perp},J_\perp,J_{\parallel})$. The dispersion relation of the conduction band electrons is assumed to be linear: $\epsilon(\vec k)=\hbar v_F (k-k_F)$, which is valid for low-energy excitations. For the convenience of the calculations we put $k_F=0$ and $\hbar=v_F=1$. Then energy and momentum are measured in the same units, and space and time also have the same units. 

The interaction in this Hamiltonian is point-like, so only s-wave scattering can occur. Therefore the model can be reduced to an effective one-dimensional model.\cite{Wil75}
We will use an ``unfolded'' picture\cite{Affleck2008} where outgoing waves correspond to $x>0$ and incoming waves to $x<0$.
 
To simplify further considerations, one can use bosonization and refermionization for this effective model.~\cite{RNM} At the special value of the coupling strength $J_{\parallel}=2-\sqrt{2}$ the $S_z$-$s^{el}_z$ interaction term between the conduction band electrons and the impurity spin vanishes and the model is reduced to a quadratic Hamiltonian:\cite{RMP_Leggett,Kehr1}
\be  H = \sum_k k :c_k^\dagger c_k: + \sum_k V\,(d^\dagger c_k +c_k^\dagger d).\ee
This is called the Toulouse limit.~\cite{Tou70} The hybridization $V$ between the fermionic operators $c_k$ and the impurity operator $d$ is proportional to the coupling strength $J_\perp$. 
The fermionic creation and annihilation operators $c^\dagger_k$, $c_k$ correspond to soliton spin excitations in the original conduction band.~\cite{RMP_Leggett} 
Normal ordering of the operators is denoted by colons~$(:\: :)$. The spin of the conduction band electrons at position $x$ can
be shown to be given by
\be s_z(x)=:c^{\dagger}(x)c(x):-1/2 \ee
where we have neglected a quickly oscillating contribution proportional to $\exp(2k_F x)$, which cannot easily be
described using bosonization techniques. In the language of Ref.~\onlinecite{Saleur2008} we are only concerned with the uniform
part of the spin susceptibility in this paper, and not the superimposed Friedel oscillations.
The spin of the Kondo impurity is
\be S_z = d^{\dagger}d-1/2 \label{impurity}.\ee

In the sequel we are interested in the situation where the interaction between the impurity and the conduction
band is switched on instaneously at $t=0$:
\be \label{ham} H = \sum_k k :c_k^\dagger c_k: + \theta(t)\sum_k V\,(d^\dagger c_k +c_k^\dagger d).\ee
We assume that for $t<0$ the new $c$-fermions are in their ground state, and the impurity spin is up before the quench. 
We denote this state with $\mid \psi_{NEQ} \rangle$. 
It is a non-stationary state for the Hamiltonian~(\ref{ham}) for $t>0$. The creation and annihilation operators have the
following expectation values in this state
\bea \label{initial}\begin{aligned}
     \langle\psi_{NEQ} \mid  d^{\dagger} d \mid \psi_{NEQ} \rangle &=&1,\\ 
\langle\psi_{NEQ} \mid d d^{\dagger} \mid \psi_{NEQ} \rangle &=&0, \\
\langle\psi_{NEQ} \mid  c_k^{\dagger} c_{k^{\prime}} \mid \psi_{NEQ} \rangle &=&n_k(\beta) \delta_{kk^{\prime}}, \\
\langle\psi_{NEQ} \mid c_k c_{k^{\prime}}^{\dagger} \mid \psi_{NEQ} \rangle &=& (1-n_k(\beta)) \delta_{kk^{\prime}}
\end{aligned}
\eea
with the thermal distribution function for the electrons at temperature $T\equiv1/\beta$ at zero chemical potential:
\be n_k(\beta)=\frac{1}{1+e^{\beta k}}.\ee

Starting from time $t=0$,
 that is the moment of coupling the impurity to the system, the conduction electrons feel the perturbation. The response to the perturbation is expressed via the time-dependent anticommutator/commutator of the Kondo spin and  the spin of the conduction electron:
\be C_{\pm}(x,t,t_w)=\langle \psi_{NEQ} \mid [S_z(t_w), s_z(x,t+t_w)]_{\pm} \mid \psi_{NEQ} \rangle \label{corr}\ee
where $[\mbox{ },\mbox{ }]_{\pm}$ denotes the anticommutator/commutator. $t_w$ is the elapsed time from the moment of turning on  the interaction between the impurity and the conduction band and the first measurement (waiting time), and $t$ is the time difference 
between the first and second spin measurement.

When an infinitesimally small magnetic field in the $z$-direction couples to the Kondo impurity at time $t_w$:
\be H_{\delta\mathcal{B}}(\tau)=H+\delta\mathcal{B}S_z\delta(\tau-t_w),\ee
then the response of the conduction band electron at position $x$ at a later time $t$ is given by the commutator
\bea &&\langle\psi_{NEQ} |s_z(x,t+t_w) |\psi_{NEQ}\rangle_{\delta\mathcal{B}} = \label{linresp} \\
&=&\langle \psi_{NEQ}|s_z(x,t+t_w) |\psi_{NEQ}\rangle +  \nonumber\\ &+&i\theta(t)\delta\mathcal{B}\langle \psi_{NEQ} | [S_z(t_w), s_z(x,t_w+t)]_{-} | \psi_{NEQ} \rangle. \nonumber\eea  
Expression~(\ref{linresp}) can be derived in the same manner as the usual linear response formula, see for example Ref.~\onlinecite{Alt}. Here we just give a brief reminder of this derivation: Let us denote the evolution operator of the system after the quench by $U(t)$, and the evolution operator with the infinitesimally small magnetic field switched on after time $t_w$ by $U_{\delta\mathcal{B}}(t)$. The difference between these two operators is $\delta U_{\delta\mathcal{B}}(t)=U^{-1}(t)U_{\delta\mathcal{B}}(t)$. It is easy to show that 
\be \frac{d \delta U_{\delta\mathcal{B}}(t)}{dt}=-i\delta\mathcal{B}(t)S_z(t)\delta U_{\delta\mathcal{B}}(t).\label{evolution}\ee 
The expectation value of the conduction band electron spin after switching on the magnetic field is
\bea &&\langle\psi_{NEQ}|s_{z}(x,t+t_w)|\psi_{NEQ} \rangle_{\delta\mathcal{B}}=\nonumber \\
&=&\langle\psi_{NEQ}|\delta U_{\delta\mathcal{B}}^{-1}(t) s_z(x,t+t_w) \delta U_{\delta\mathcal{B}}(t) |\psi_{NEQ} \rangle.\label{derlinresp}\eea
Plugging  $\delta\mathcal{B}(t)=\delta\mathcal{B}\,\delta(t-t_w)$ into the
equation for the evolution operator~(\ref{evolution}) and then using~(\ref{derlinresp}) yields~(\ref{linresp}).

The relation~(\ref{linresp}) gives the physical interpretation of $C_-(x,t,t_w)$:} it describes the linear response to a perturbation
acting on the Kondo spin at time~$t_w$ after the initial quench. From causality we therefore expect no response 
outside the light cone, that is for distances $x>t$. 
The anticommutator in (\ref{corr}), on the other hand, is a {\it symmetrized correlation function}. 
It does not have such a linear response interpretation for observables, but characterizes the spread of entanglement in the system. 
For fermions it is directly connected to the entanglement entropy.~\cite{ent} 

The scheme for the calculation of the commutator/anticommutator~(\ref{corr}) is the following:~\cite{Kehr}
\begin{itemize}
\item make a unitary transformation of the non-diagonal Hamiltonian~(\ref{ham}) at $t>0$ to its diagonal form;
\item evolve the operators with the diagonal Hamiltonian;
\item make the reverse unitary transformation to the initial operators.
\end{itemize}
This approach allows us to get a non-perturbative solution for the time evolution of the model under consideration. 

The detailed calculation can be found in Ref.~\onlinecite{thesis}, here we only give a concise recapitulation.
The Hamiltonian~(\ref{ham}) for $t>0$ is diagonalized by the following Bogolyubov transformation:
\be a_\epsilon=A_{\epsilon d}d + \sum_k A_{\epsilon k}c_k \label{transform}.\ee
We can write the diagonalized Hamiltonian as
\be H= \sum_\epsilon \epsilon a_\epsilon^\dagger a_\epsilon \label{quadro}\ee
with the coefficients
\be A_{\epsilon d}=\sqrt{\frac{\Delta_L}{\pi}\frac{\Delta}{\epsilon^2+\Delta^2}}, 
A_{\epsilon k}=\frac{V}{\epsilon-k}\sqrt{\frac{\Delta_L}{\pi}\frac{\Delta }{\epsilon^2+\Delta^2}}.\ee
We consider a lattice of length~$L$ such that the values of~$k$ are quantized. The difference between the neighboring momenta is denoted by $\Delta_L=2\pi/L$. The hybridization function is denoted by  $\Delta$  and defined as $\Delta(\epsilon)=\pi\sum_k V^2\delta(\epsilon-k)=V^2L/2$. From bosonization and refermionization one can identify the parameter $\Delta$ with the Kondo temperature via 
\begin{equation}
\Delta=\frac{T_K}{\pi w} \ ,
\label{TK} 
\end{equation}
where $w\approx0.42$ is the Wilson ratio.~\cite{Wil75,Aff91}

The dynamics of the operators $a_{\epsilon}$ governed by the quadratic Hamiltonian~(\ref{quadro}) is simple
\be a_{\epsilon}(t)=a_{\epsilon}(0)\exp(-i\epsilon t).\ee
We arrive at the initial operators $c(x)$ and $d$ by making the inverse transformation of~(\ref{transform}). This gives us expressions for the time evolution of the operators $c(x,t)$ and $d(t)$. 
Now we can get the expressions for the commutator and anticommutator~(\ref{corr}) by taking into account the properties of the initial state determined by~(\ref{initial}):
\begin{widetext}
\bea \label{cplus}&\langle\psi&_{NEQ} \mid [S_z(t_w), s_z(x,t_w+t)]_{+} \mid \psi_{NEQ} \rangle= \nonumber\\
&=&\begin{cases} 
0<x<t+t_w& \Delta\theta(t-x)e^{2\Delta (x-t) }-\Delta \sigma^2(x,t,t_w)    
\\
\mbox{else } &-\Delta \left [ c_{\beta}(x-t) -s_{\beta}(x-t)- e^{-\Delta t_w}(c_{\beta}(x-t-t_w)-s_{\beta}(x-t-t_w))\right]^2 
\end{cases}
\eea
and
\bea \label{cminus}\langle\psi_{NEQ} \mid [S_z(t_w), s_z(x,t_w+t)]_{-} \mid \psi_{NEQ} \rangle= 
\begin{cases} 
0<x<t& 2i \Delta\,e^{\Delta (x-t) } \sigma(x,t,t_w)    
\\
\mbox{else } &0,
\end{cases}
\eea
with the notation
\be \sigma(x,t,t_w)=c_{\beta}(x-t)+s_{\beta}(x-t) 
-e^{-\Delta t_w}(c_{\beta}(x-t-t_w)+s_{\beta}(x-t-t_w))-2e^{\Delta(x-t-t_w)}s_{\beta}(t_w)\ee
\end{widetext}
where the functions $c_{\beta}(x)$ and $s_{\beta}(x)$ are determined as
\bea 
\label{cbeta} c_{\beta}(x)= \frac{1}{\pi \Delta^2} \int_0^{\infty} dk \frac{n_k(\beta) k\cos(kt)}{1+(k/\Delta)^2}, \\
\label{sbeta} s_{\beta}(x)= \frac{1}{\pi \Delta} \int_0^{\infty} dk \frac{n_k(\beta) \sin(kt)}{1+(k/\Delta)^2}.
\eea
Expressions (\ref{cplus}) and (\ref{cminus}) are the main formula of our paper and contain exact results about the full spatiotemporal buildup of the
Kondo correlations. In the following section we will analyze their physical interpretation.  Let us already mention that for infinite waiting time $t_w\rightarrow\infty$ one recovers the equilibrium behavior:~\cite{thesis,Kehr1} 
\be 
\lim_{t_w\rightarrow\infty} C_{\pm}(x,t,t_w) = C^{eq}_{\pm}(x,t)
\ee
Clearly such an approach to equilibrium is expected for an impurity model. 
Specifically, the equilibrium decay of the correlations in the strong coupling limit~\cite{Aff95} is proportional to $1/x^2$. 

\section{Correlation functions}

\begin{figure}[tb] 
\includegraphics[width=0.75\linewidth]{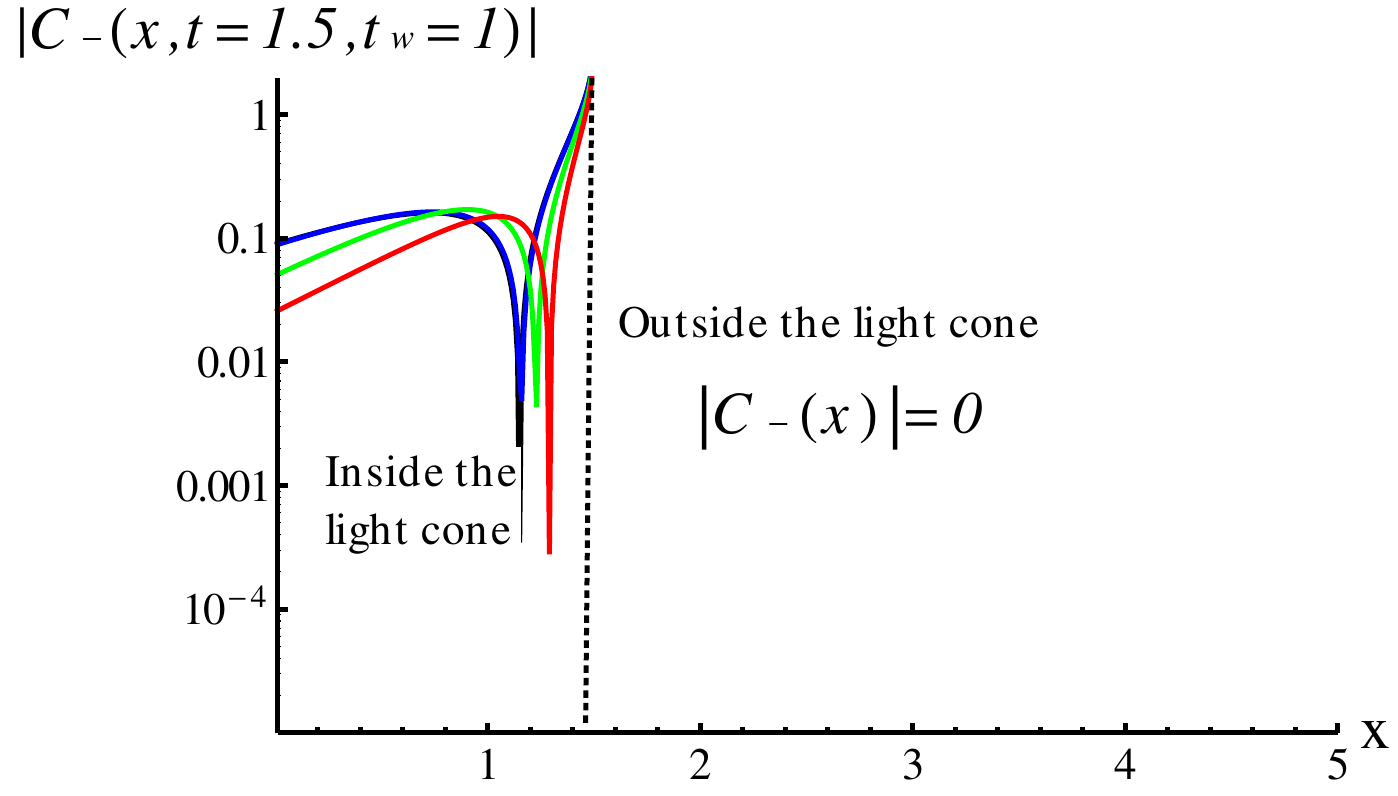}
\caption{\label{cone2dminus}
The dependence of the commutator on the distance from the impurity at fixed waiting time $t_w=1$ and time difference between
the measurements $t=1.5$ for different temperatures. $\mid C_{-}(x,t=1.5,t_w=1)\mid$ is plotted for $\beta=\infty$ (black line), $\beta=5$ (blue line), $\beta=1$ (green line), $\beta=0.5$ (red line). All quantities are measured in units of $\Delta$ corresponding to the
Kondo temperature according to (\ref{TK}). Notice that the commutator is strictly zero outside the light cone. 
}
\end{figure}

\begin{figure}[ht!] 
A\includegraphics[width=0.75\linewidth]{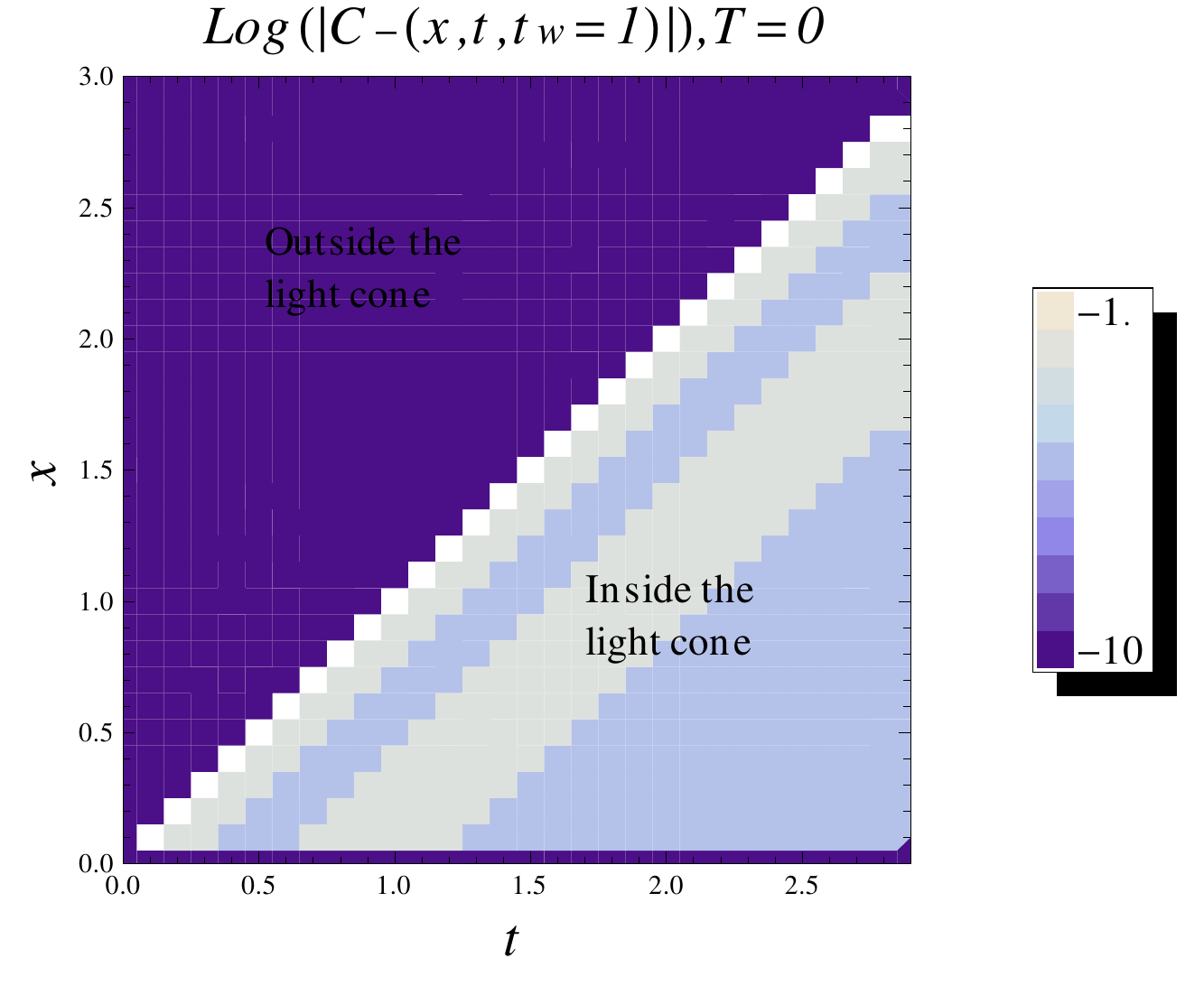}
B\includegraphics[width=0.75\linewidth]{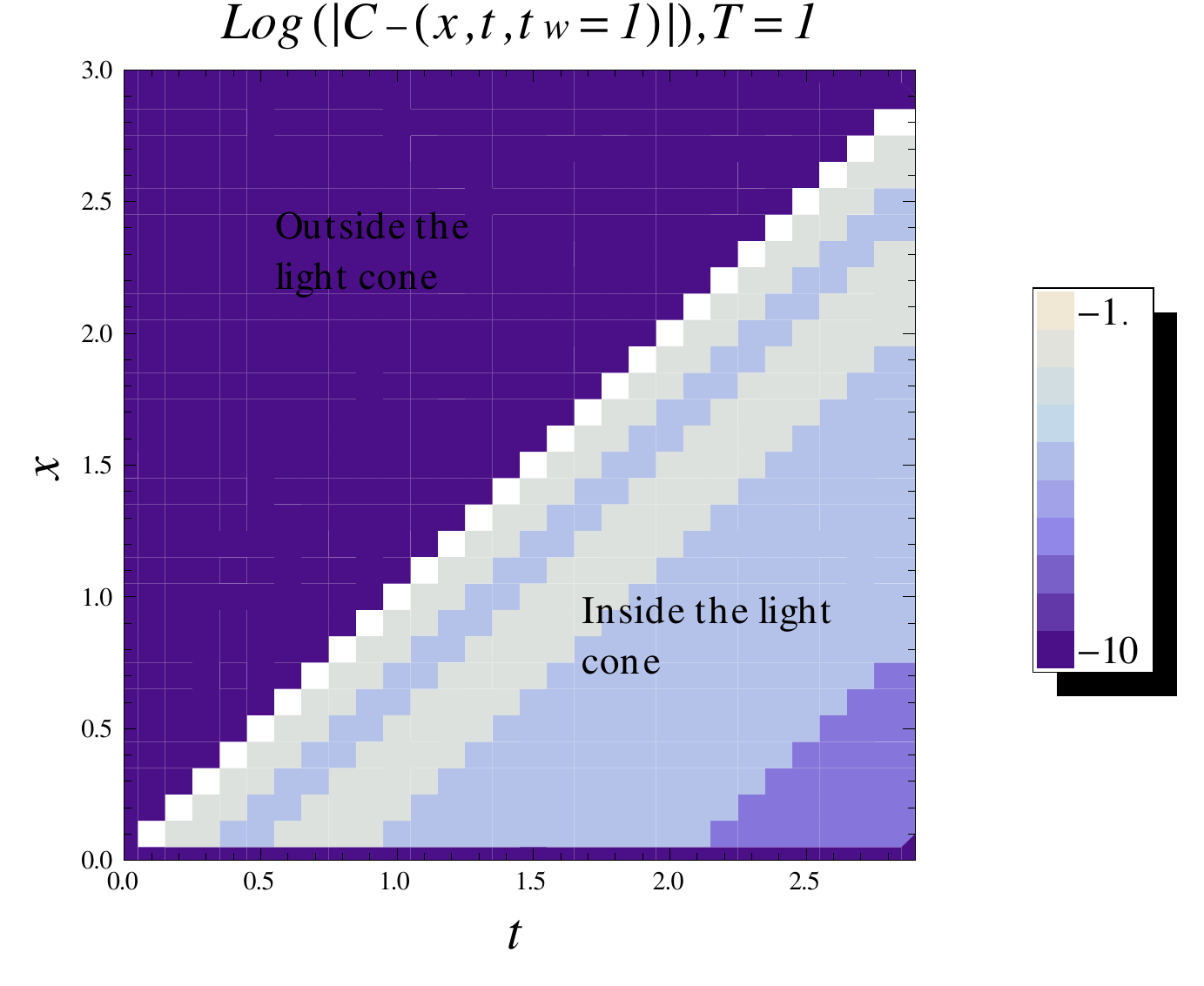}
\caption{\label{cone3dminus}
The spatiotemporal behavior of the response function (commutator), $\mid C_{-}(x,t, t_w=1)\mid$, (A) at zero and (B) non-zero temperature ($\Delta=1$). We clearly see the light cone: the commutator vanishes exactly outside the light cone due to causality. 
}
\end{figure}

\begin{figure}[ht!] 
\includegraphics[width=0.75\linewidth]{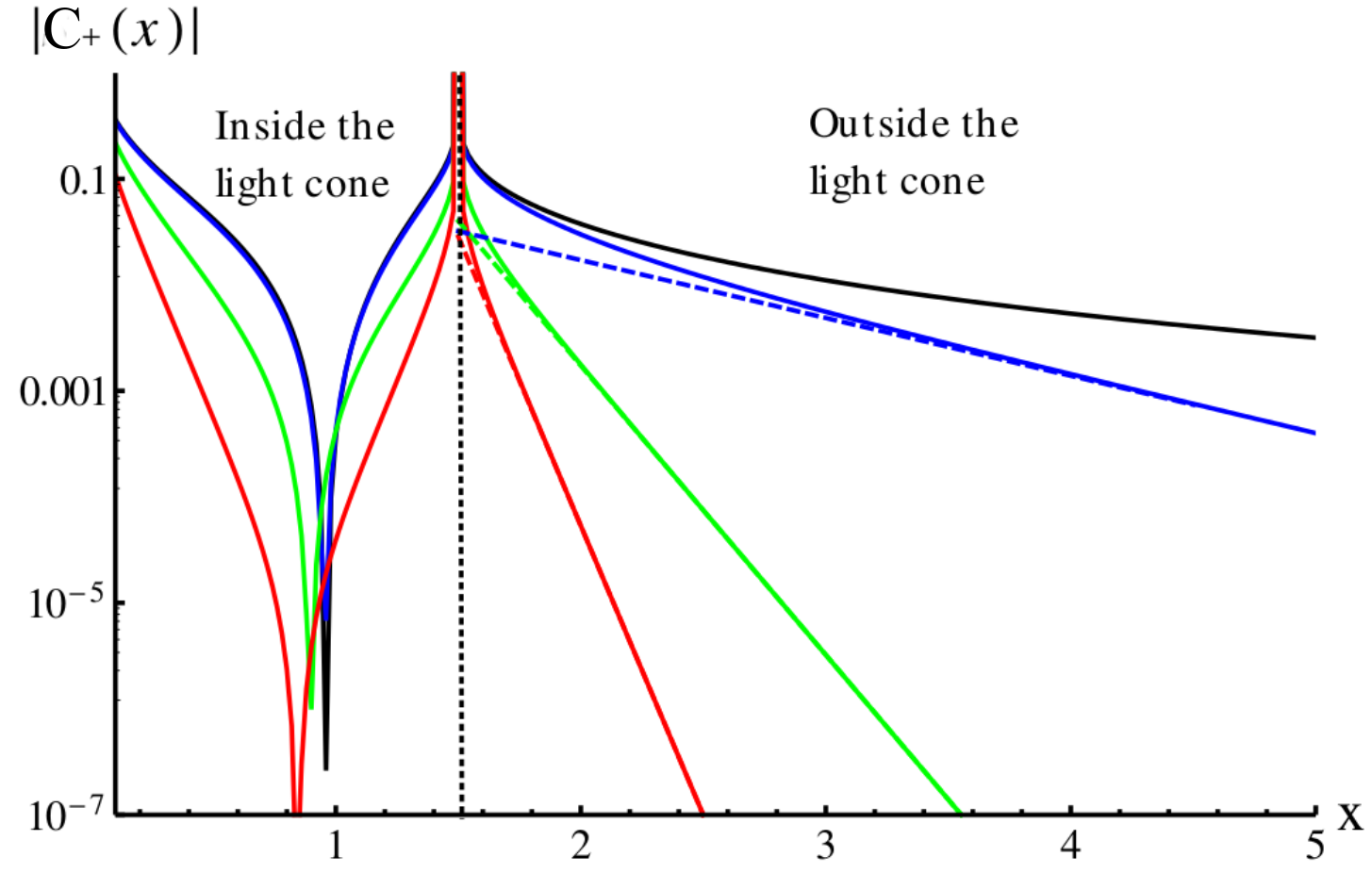}
\caption{\label{cone2dplus}
The dependence of the equal time anticommutator, $\mid C_{+}(x,t=0,t_w=1.5)\mid$,
on the distance from the impurity for fixed waiting time $t_w=1.5$  for different temperatures: $\beta=\infty$ (black line), $\beta=5$ (blue line), $\beta=1$ (green line), $\beta=0.5$ (red line) (always $\Delta=1$). One clearly sees a light cone like peak at $x=t_w$.
For zero temperature ($\beta=\infty$) the correlations decay with a power law outside the light cone. For high temperatures ($\beta=1$, $\beta=0.5$) the decay is exponential consistent with the asymptotic expression (\ref{Tn0}). For intermediate temperature, $\beta=5$, there is a crossover between a power law decay (close to the light cone) and an exponential decay (far away from the light cone).
}
\end{figure}

\begin{figure}[ht!] 
A\includegraphics[width=0.75\linewidth]{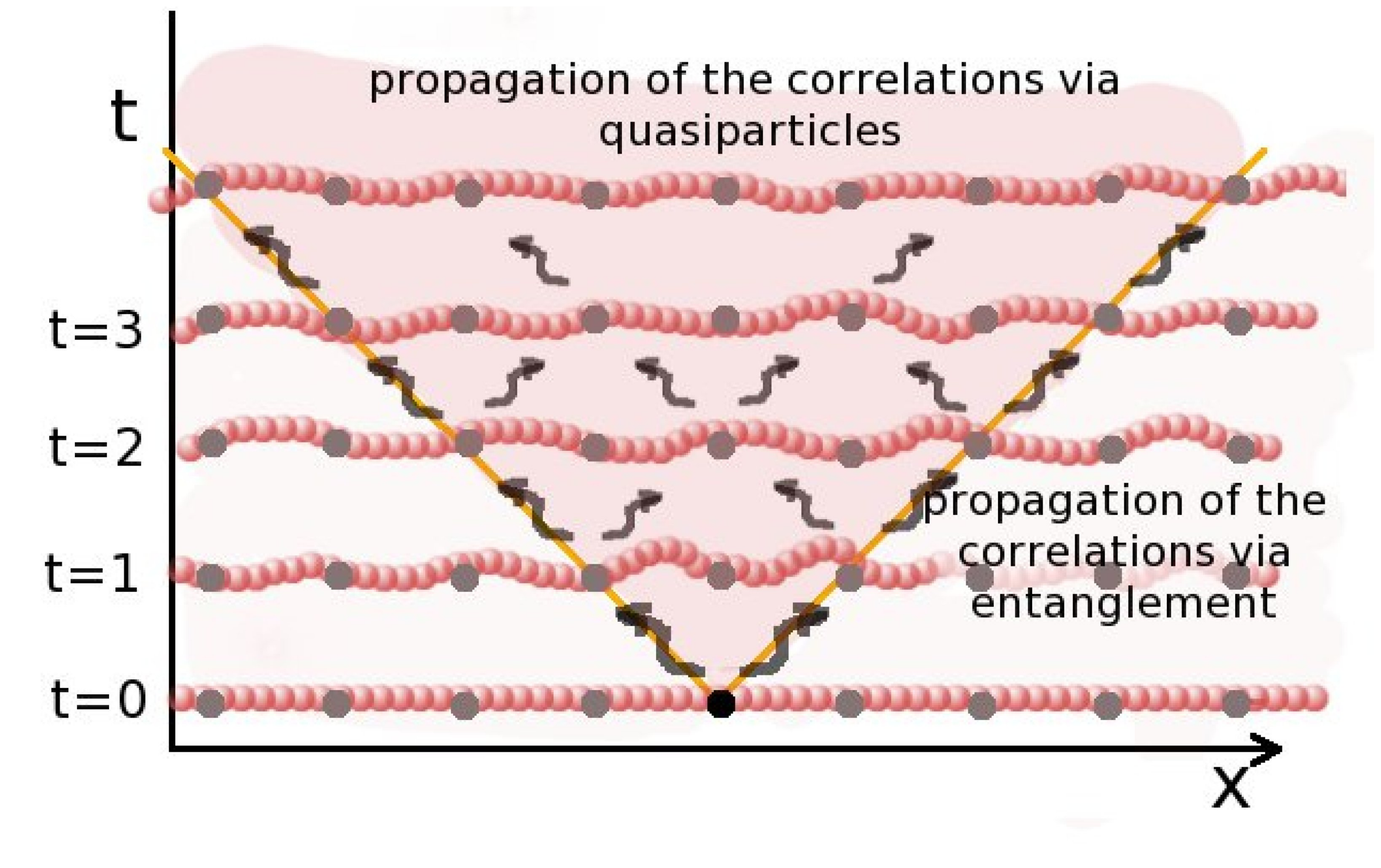}
B\includegraphics[width=0.75\linewidth]{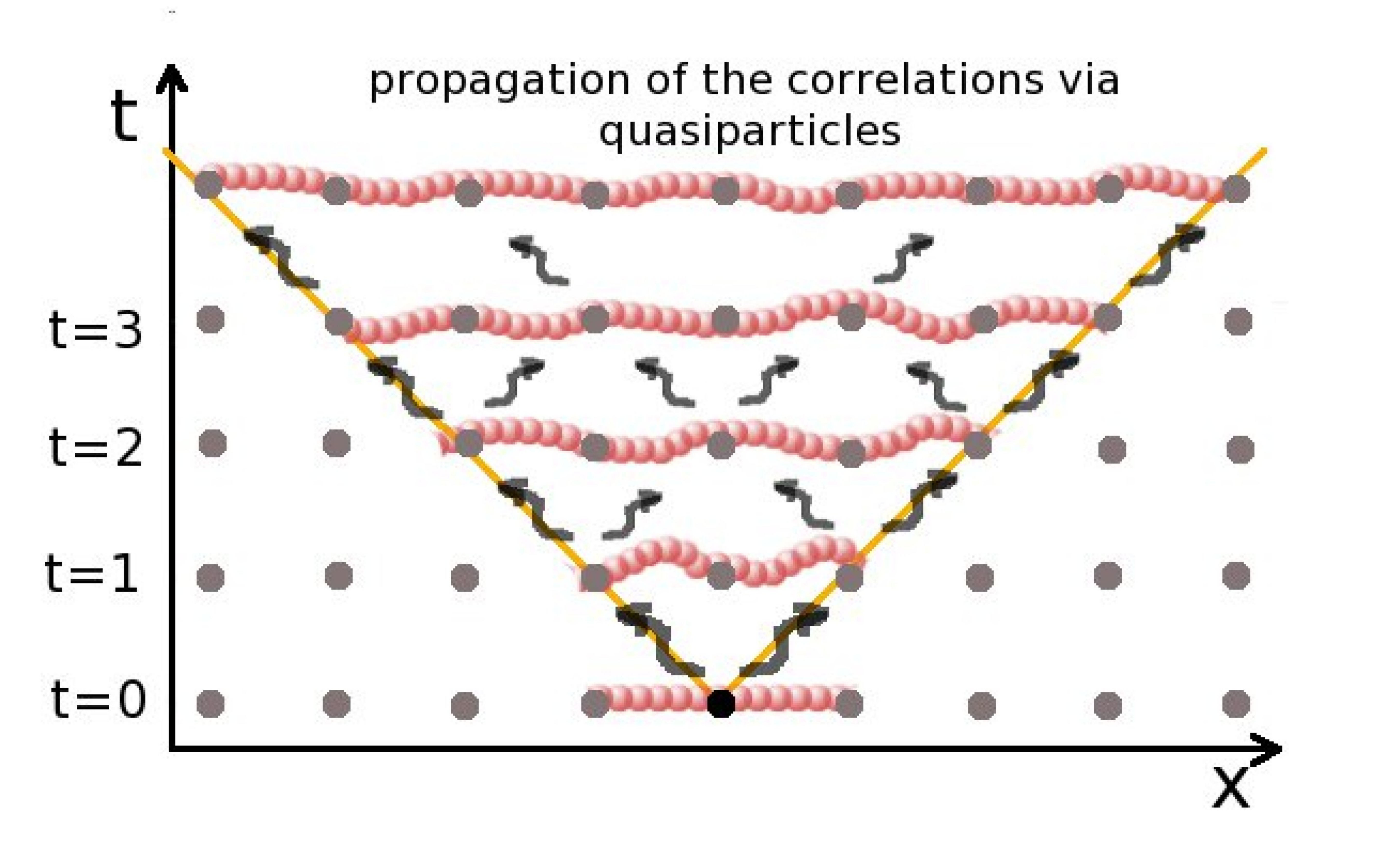}
\caption{\label{spread}
Schematic representation of the spread of correlations for a state with initial long range entanglement (A) vs. one with short range entanglement (B). Correlations spread via quasiparticles inside the light cone, while outside the light cone correlations build up via the initial entanglement. Solid circles are lattice sites, smeared circles denote the entanglement between the lattice sites.  
}
\end{figure}

\begin{figure}[ht!] 
A\includegraphics[width=0.75\linewidth]{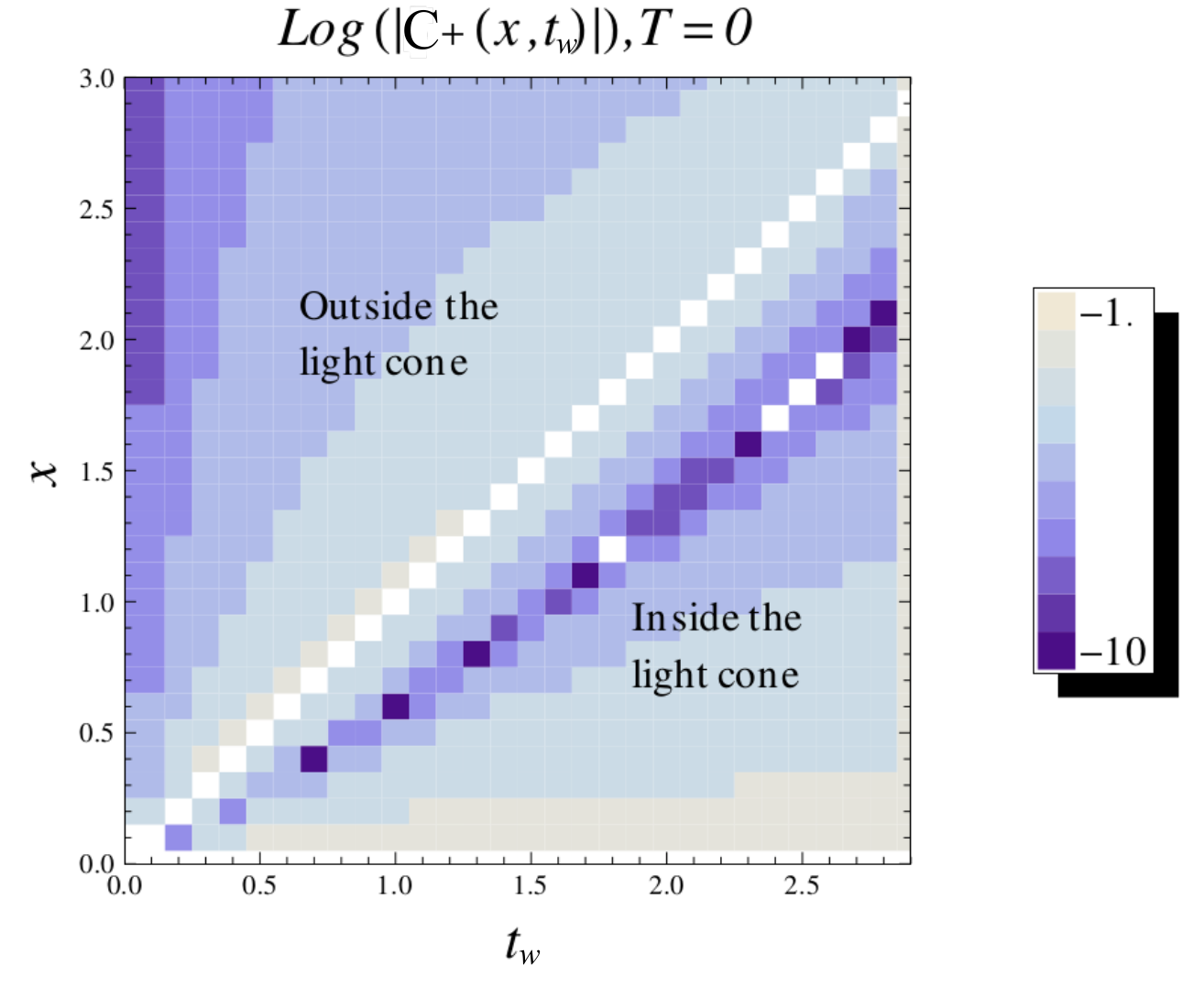}
B\includegraphics[width=0.75\linewidth]{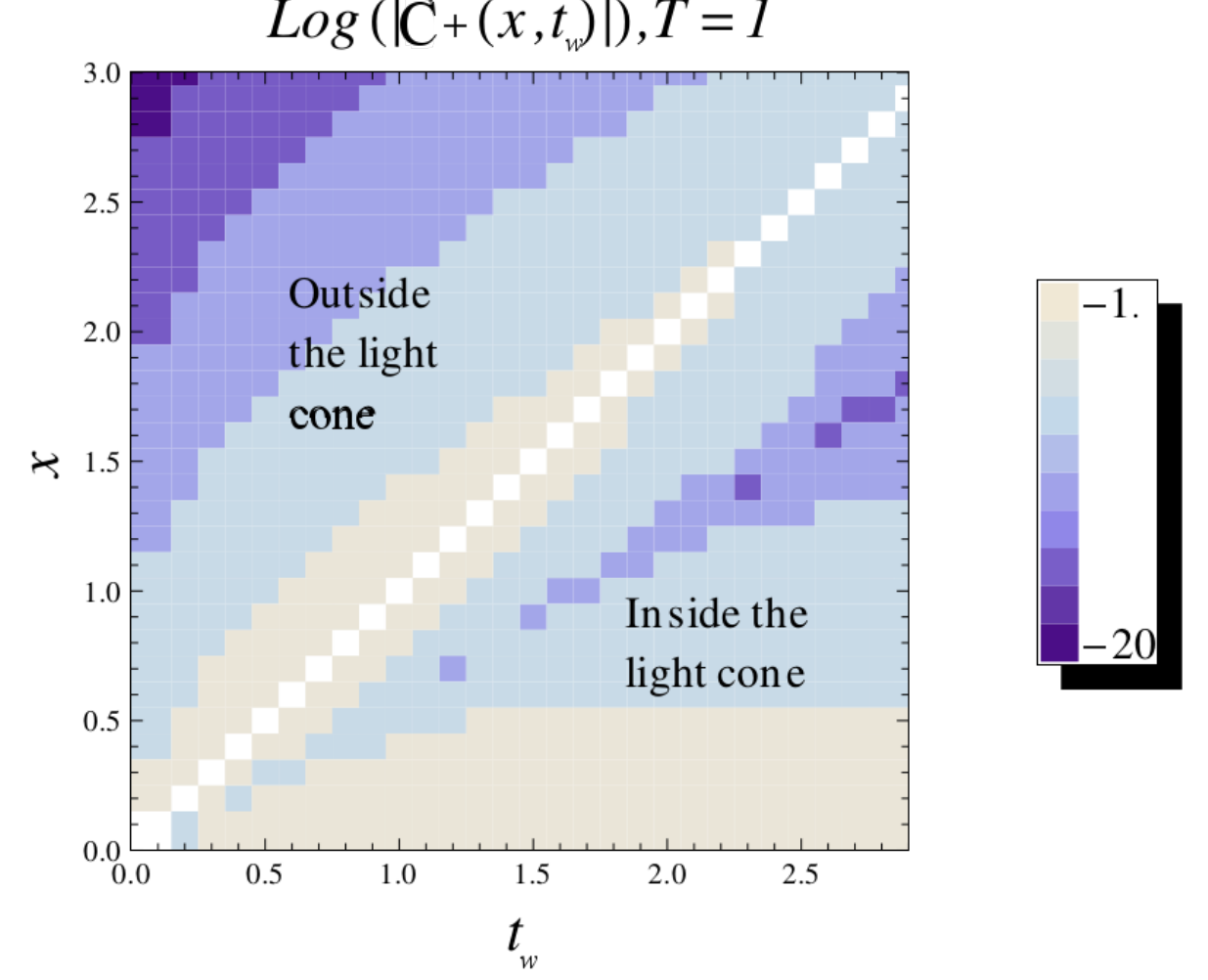}
\caption{\label{cone3dplus}
The spatiotemporal behavior of the correlation function (anticommutator), $\mid C_{+}(x,t=0,t_w)\mid$, at zero (A) and non-zero (B) temperature ($\Delta=1$). One clearly observes the light cone structure for $x=t$. Outside the light cone the decay of the correlation function is power law like at zero temperature, while at non-zero temperature $C_{+}$ decays exponentially.
}
\end{figure}

Let us analyze the expressions for the commutator and anticommutator (\ref{corr}). 
The commutator, $C_{-}(x,t,t_w)$, is zero for $x>t$ -- outside the light cone, see Eq.~(\ref{cminus}) and Fig.~\ref{cone2dminus}. As discussed before this is the  expected result since the commutator has a linear response interpretation. The conduction band electrons are effectively a relativistic system and causality leads to a finite propagation speed of the perturbation.  
Let us also note that the commutator at $t_w=0$ is zero since a perturbation at $t_w=0$ 
will not affect the system, which is initially prepared in an eigenstate of~$S_z$.  
The further growth of the response is proportional to $t_w$ at small waiting times, $t_w\ll 1$. This follows directly from the expressions~(\ref{cminus}).
We show the full spatiotemporal dependence of the commutator in Fig.~\ref{cone3dminus}.

Now let us look at the {\it equal time} anticommutator, $C_+(x,t=0,t_w)$.
This describes the correlations between the impurity spin and some spatially separated electron spin at {\it the same moment of time} ($t=0$). The commutator was identically zero for such spatiotemporal configurations due to causality. However, the anticommutator is non-zero, compare Figs.~\ref{cone2dplus} and~\ref{cone3dplus}.
Notice that the anticommutator is not measurable in a single experiment (it does not have a linear response interpretation), hence there is no contradiction with the causality principle.  $C_{+}$ can be determined as a statistical property, for example, by weak measurements of the conduction band spin at point $x$ and the impurity spin.

\begin{figure}[htb] 
\includegraphics[width=0.75\linewidth]{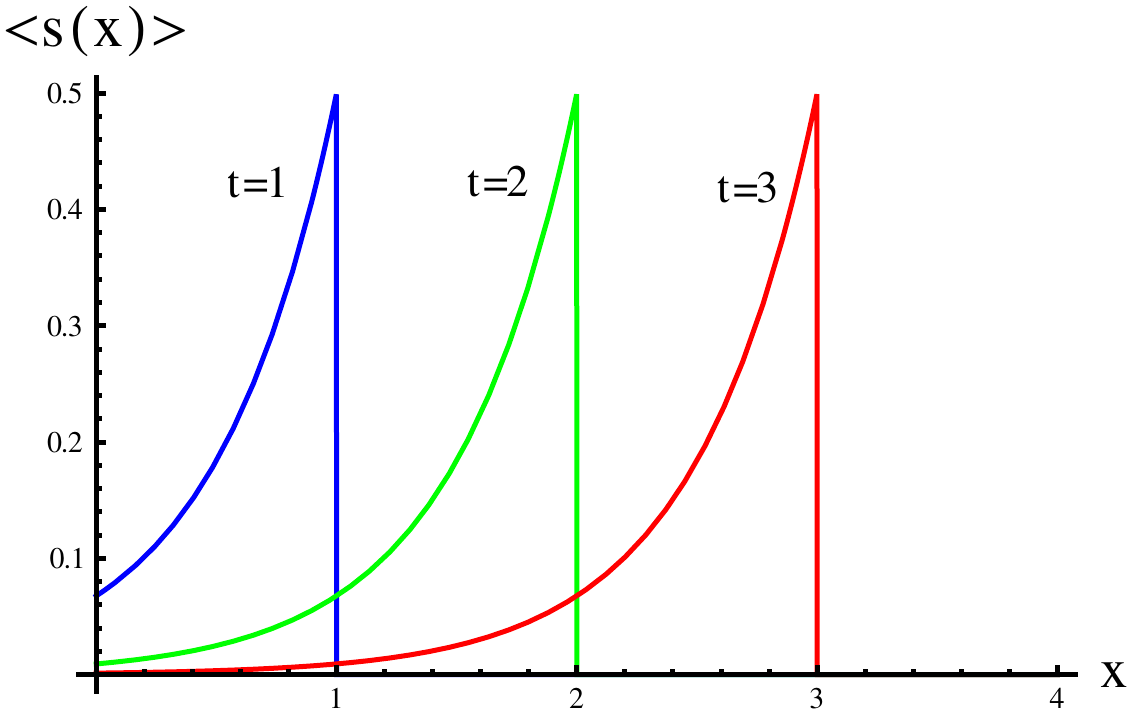}
\caption{\label{s1d}{
The expectation value of the conduction band electron spin~(\ref{s1}) at different times clearly exhibits the light-cone effect.}}
\end{figure}

Notice that now a light cone can be seen at $x=t_w$: $C_{+}(x,t=0,t_w)$ has a peak (actually a divergence) for 
$x=t_w$. The divergence is nonphysical. We ascribe its presence to the assumption that the coupling between the impurity and the conduction band does not depend on the momentum. In a realistic system this coupling decays for large values of momentum, leading to a finite value of $C_{+}$ at $x=t_w$.
The correlations are largest on the light cone itself since this point just represents the ballistic spin transport away from the impurity. 
One can see this nicely from  
the expectation value of the conduction band electron spin after the quench according to (\ref{corr})
\bea\label{s1}
 \langle \psi_{NEQ}|s_z(x,t)|\psi_{NEQ}\rangle=C_+(x,t,t_w=0)= \\
 =\begin{cases} 0<x<t & \frac{1}{2}\,e^{2\Delta(x-t)} \\
  \mbox{else } & 0
 \end{cases}
\eea
The transport of spin to spatial infinity is represented in
Fig.~\ref{s1d} where the value $\langle s_z(x)\rangle$ is shown for different times. 
This spin transport to infinity is essential for the formation of the Kondo singlet ground state at large times.

Outside the light cone (for $x>t_w$) the correlations decay either with a power law (zero temperature), or exponentially (non-zero temperature). One can interpret 
the light cone in $C_{+}(x,t=0,t_w)$ as showing the spread of correlations in the system: Initially there are no correlations
between the impurity spin and the conduction band (\ref{initialstate}), but after the quench
 quasiparticles carry the information about the perturbation to other parts of the system. 
 
 The non-vanishing correlations outside the light cone for $t_w>0$ are ascribed to the initial entanglement of the electrons in the bath. Notice that while $|{\rm FS}\rangle$ is not entangled in the momentum representation, in the coordinate representation it is entangled. Inside the light cone the correlations spread via quasiparticles, while the tails outside the light cone result from the initial entanglement in the system. This is schematically depicted in Fig.~\ref{spread}. Essentially, for any nonzero waiting time
 the impurity spin becomes entangled with the conduction band spin localized at the impurity site, which in turn is already
 entangled with conduction band spins far away due to the structure of $|{\rm FS}\rangle$. This immediately (for infinitesimal $t_w>0$) leads to
 entanglement between the impurity spin and far away conduction band spins resulting in the tails outside the light
 cone, which therefore do not violate any causality condition.
 Similar behavior has also been seen in Ref.~\onlinecite{SCqubits}.

The decay of $C_{+}(x,t=0, t_w)$ outside the light cone is well described by the following asymptotic expressions (see appendix~A for the derivation):
For $\beta\Delta\ll1$ one finds
\begin{widetext}
\be \label{Tn0} C_{+}(x,t=0,t_w)\approx \Delta \frac{\left[\tfrac{2}{\pi} \left(1+\tfrac{\beta \Delta}{\pi}\right)\right]^2}{\left[1-\left(\tfrac{\beta\Delta}{\pi}\right)^2\right]^2}\exp\left(-\frac{2(x-t_w)\pi}{\beta} \right) \left(\exp\left(-t_w(\pi/\beta) \right) -\exp(-t_w\Delta)\right)^2
+O\left(\exp\left(-\frac{3t_w\pi}{\beta} \right)\right)\ee
\end{widetext}
which shows an exponential decay as a function of distance. 
On the other hand, at zero temperature the decay is power-law-like proportional to~$x^{-2}$
\be C_{+}(x,t=0,t_w)\approx \frac{1}{\pi^2\Delta}\left( -\frac{e^{-2t_w\Delta}}{(x-t_w)^2} + \frac{2e^{-t_w\Delta}}{(x-t_w)x}-\frac{1}{x^2}\right)\label{T0}.\ee
The different decay behavior comes from the different behavior of the correlations in a Fermi gas: at zero temperature the correlations in the ground state decay as a power law, while for non-zero temperature the correlations are  decaying exponentially. So effectively the temperature reduces the entanglement of the ground state. 
Coming back to Fig.~\ref{cone2dplus}, we can see the absolute value of the anticommutator $C_{+}(x,t=0,t_w=\mbox{const.})$  for different temperatures.
We clearly see that the behavior outside the light cone indeed approaches the asymptotics given by Eqs.~(\ref{Tn0}) and (\ref{T0}).

\begin{figure}[htb] 
A\includegraphics[width=0.75\linewidth]{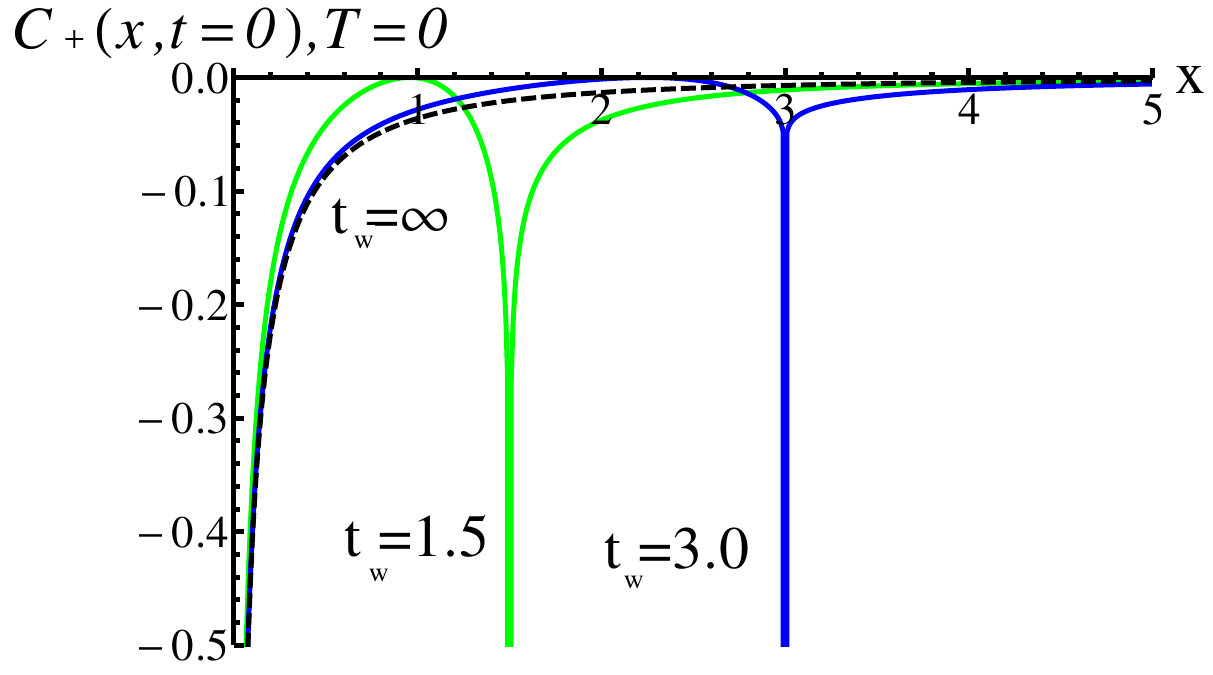}
B\includegraphics[width=0.75\linewidth]{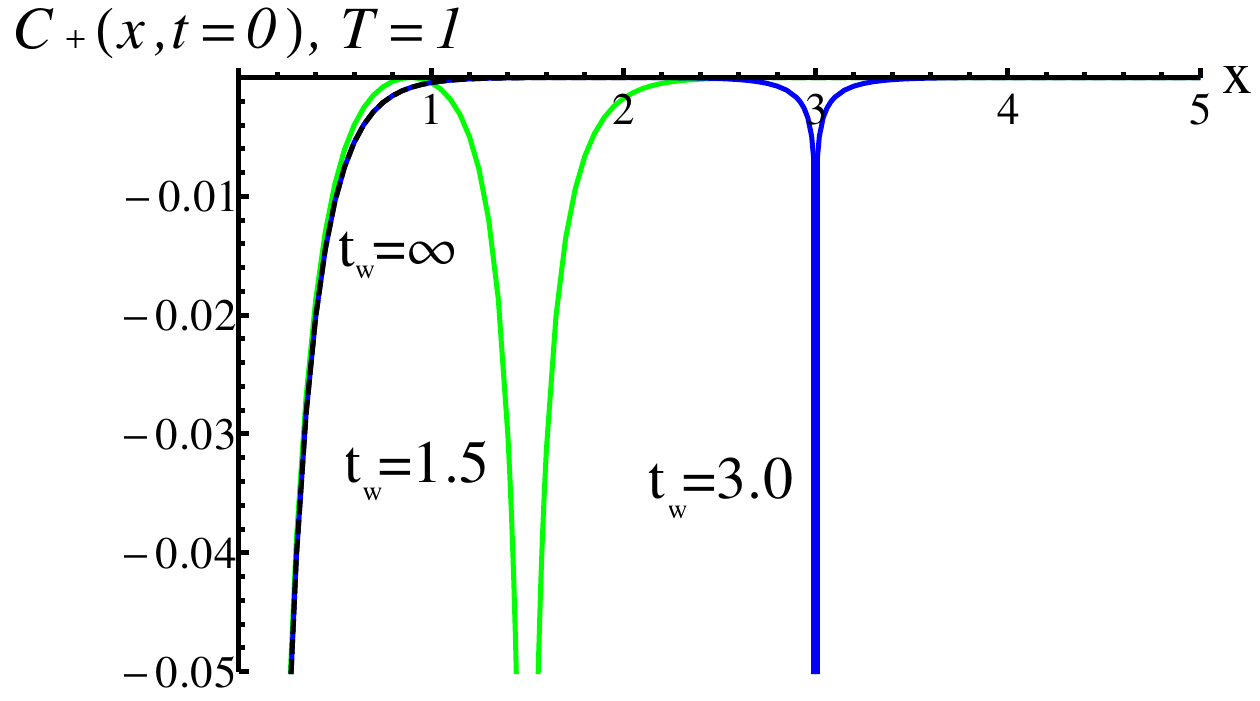}
\caption{\label{conv}
Buildup of the Kondo screening cloud: $C_+(x,t)$ for different waiting times,  $t_w=1.5$ (blue), $t_w=3.0$ (green), $t_w=\infty$
(black dashed).}
\end{figure}

In Fig.~\ref{conv} the equal-time correlation function for different waiting times is depicted on a logarithmic scale. 
The dip in the plots of $\mid C_{+}(x,t)\mid$ represents a zero of the correlation function. Close to the impurity there are
antiferromagnetic correlations indicating the buildup of the Kondo screening cloud as depicted in Fig.~\ref{conv}. One sees
how one approaches the
equilibrium Kondo screening cloud for $t_w=\infty$.

\begin{figure}[htb] 
A\includegraphics[width=0.75\linewidth]{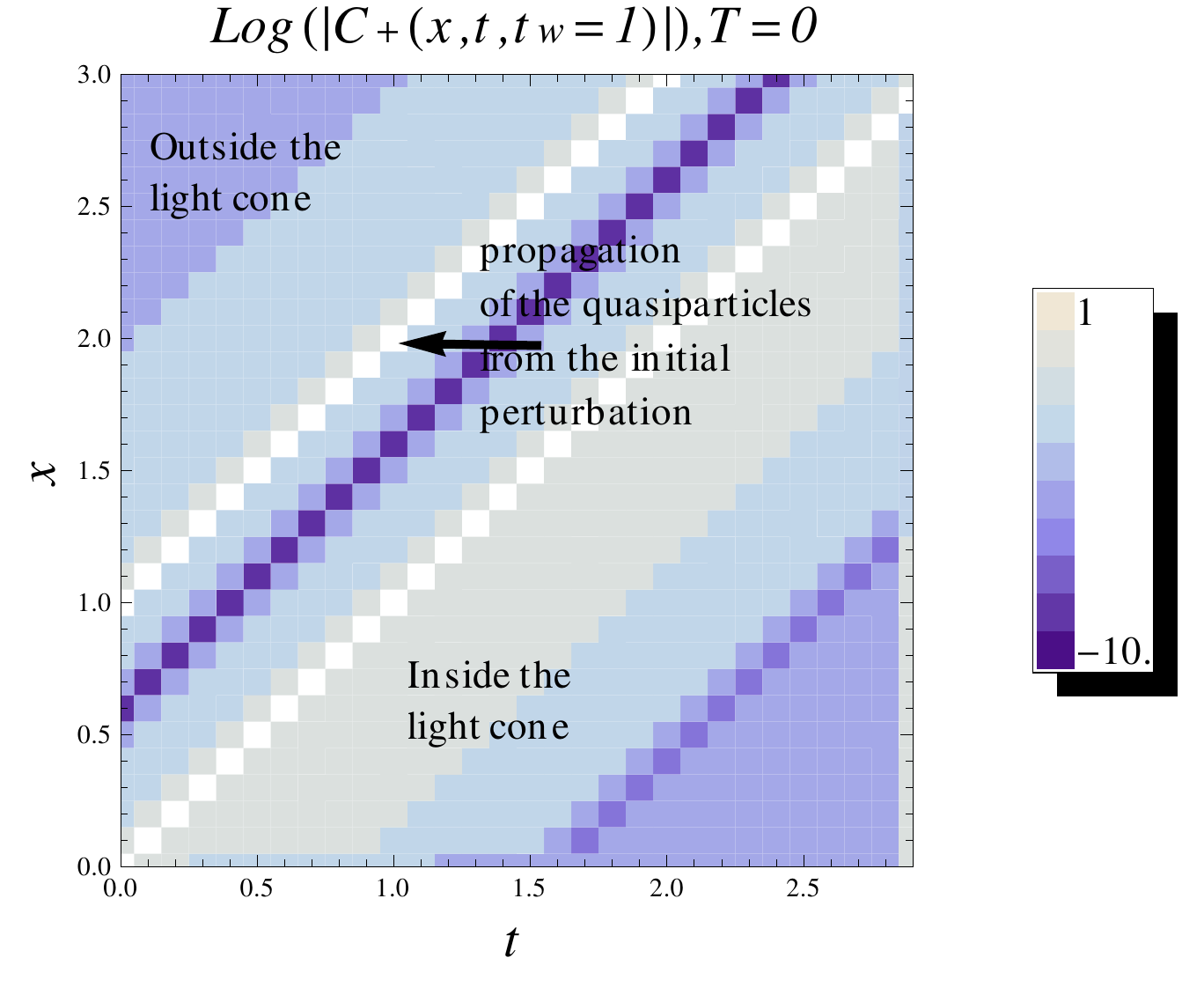}
B\includegraphics[width=0.75\linewidth]{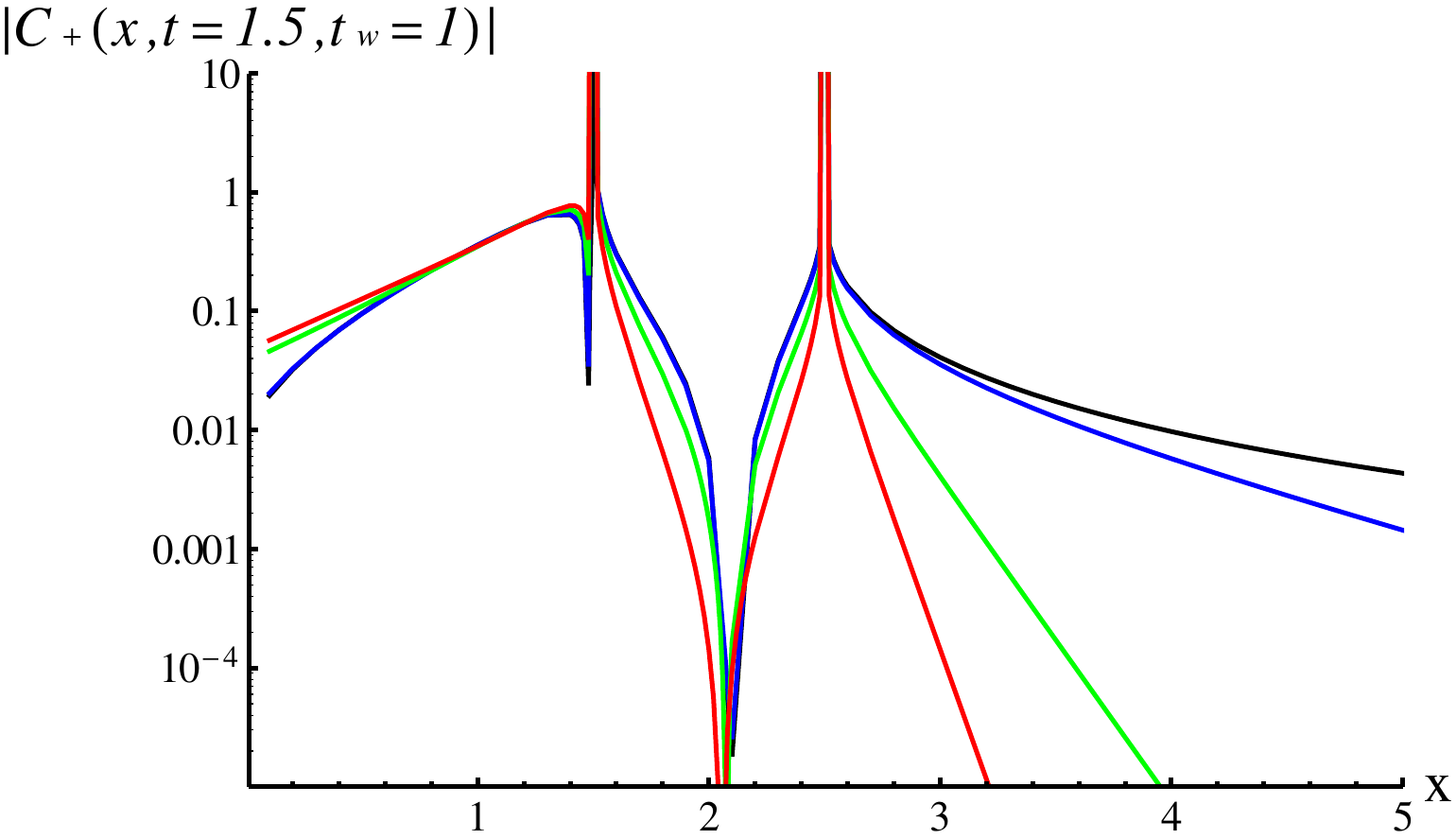}
\caption{\label{twconst}
The non-equal time correlation function (depicted for $t_w=1$) exhibits a double light-cone structure: One light cone starts from $x=1, t=0$ and represents the spread of the correlations via the quasiparticles coming from the initial coupling of the impurity spin to the conduction band, that is the
spin transport to infinity. The second light cone starts at $x=0, t=0$ and shows the correlations due to the first spin
measurement at $t_w=1$. The color scheme is the same as in Fig.~\ref{cone2dplus}.
}
\end{figure}

Let us finally look at the spatiotemporal structure of the non-equal time correlation function (that is the anticommutator) in Fig.~\ref{twconst}.  One finds a double light-cone structure:
One light cone starts from $x=t_w,t=0$ and represents the spread of the correlations via the quasiparticles coming from the initial coupling of the impurity spin to the conduction band, that is the
spin transport to infinity. The second light cone starts at $x=0, t=0$ and shows the correlations due to the first spin
measurement at~$t_w$.

\section{Conclusions}

We have investigated the time and space dependence of the spin correlation functions at the exactly solvable Toulouse point of the Kondo model. Using bosonization and refermionization techniques, we have been able to obtain exact analytical results for these
correlation functions starting from an initially unentangled product state (\ref{initialstate}). 

In our results we have seen a clear difference between the commutator (susceptibility) and the anticommutator 
of the impurity spin and the conduction electron spin.
The commutator is related to the response to a perturbation (\ref{linresp}) and vanishes exactly outside the effective light cone. 
On the other hand, while the equal-time correlator (anticommutator) also exhibits a light cone structure, it also develops tails outside the light cone. 
The light cone itself develops due to the propagation of quasiparticles  originating from the initial perturbation. Outside the light cone, the entanglement which is already present in the uncoupled Fermi gas assists the buildup of correlations,  see Fig.~\ref{spread}.  Indeed, temperature decreases the initial entanglement in the system, therefore for $T>0$ the tails decay exponentially as opposed to algebraically $\propto x^{-2}$ at zero temperature. 

Notice that the buildup of correlations outside the light cone does not contradict causality. The resolution of the apparent paradox -- there are non-zero correlations outside the light cone -- comes from the understanding of the measurement of the correlation function. The correlation function is a statistical property of the system. To determine it, one needs to  know simultaneously both the values of the impurity spin and the spin of the electron in the conduction band. Non-zero correlations outside the light cone can exist, but this does not imply the spread of information with superluminal velocities. 
Similar paradoxes appeared and were resolved in other quantum systems, both entangled (for example, the famous Einstein-Podolsky-Rosen paradox~\cite{EPR}) and non-entangled (for example, the Hartman effect -- apparent superluminal propagation under the tunnel barrier~\cite{Hart,Winful}).


Let us finally discuss the connection between our results and Lieb-Robinson bounds.~\cite{Nac} 
 Initially, the Lieb-Robinson bounds were formulated for the non-equal time commutator of two spins in a lattice model: this commutator is exponentially small outside the light cone.~\cite{LiebRobinson}  One can use this to show that even in a non-relativistic theory
the speed of the propagation of information is limited by the Lieb-Robinson bound.~\cite{Bra06,Schuch}
In our model this statement is obvious since we have an effectively relativistic theory and 
the commutator is exactly zero outside the light cone. The fact that the equal time correlation function (anticommutator) 
at zero temperature has an
algebraic tail outside the light cone in our model is not in contradiction to Lieb-Robinson bounds: Without additional 
assumptions these do not
make statements about equal time correlation functions. Generalizations of the Lieb-Robinson bounds that
are applicable to our situation make the assumption of short-range entanglement in the initial state.~\cite{Bra06}
However, precisely this condition is violated in the Fermi gas part of our initial state~(\ref{initialstate}), which in turn
is responsible for the algebraic tail found in (\ref{T0}).


An interesting issue to be addressed in future work is a comparison of the spread of entanglement (which can be connected to the anticommutator for a fermionic system) in a relativistic and non-relativistic conduction band. 

\section*{Acknowledgments}
We are grateful for helpful discussions with John Cardy, Mihailo Cubrovic, Jens Eisert, Corinna Kollath and Salvatore Manmana. 
\appendix

\section{Asymptotic estimates for $\beta\ll t$}
\begin{figure}[htb] 
\includegraphics[width=0.75\linewidth]{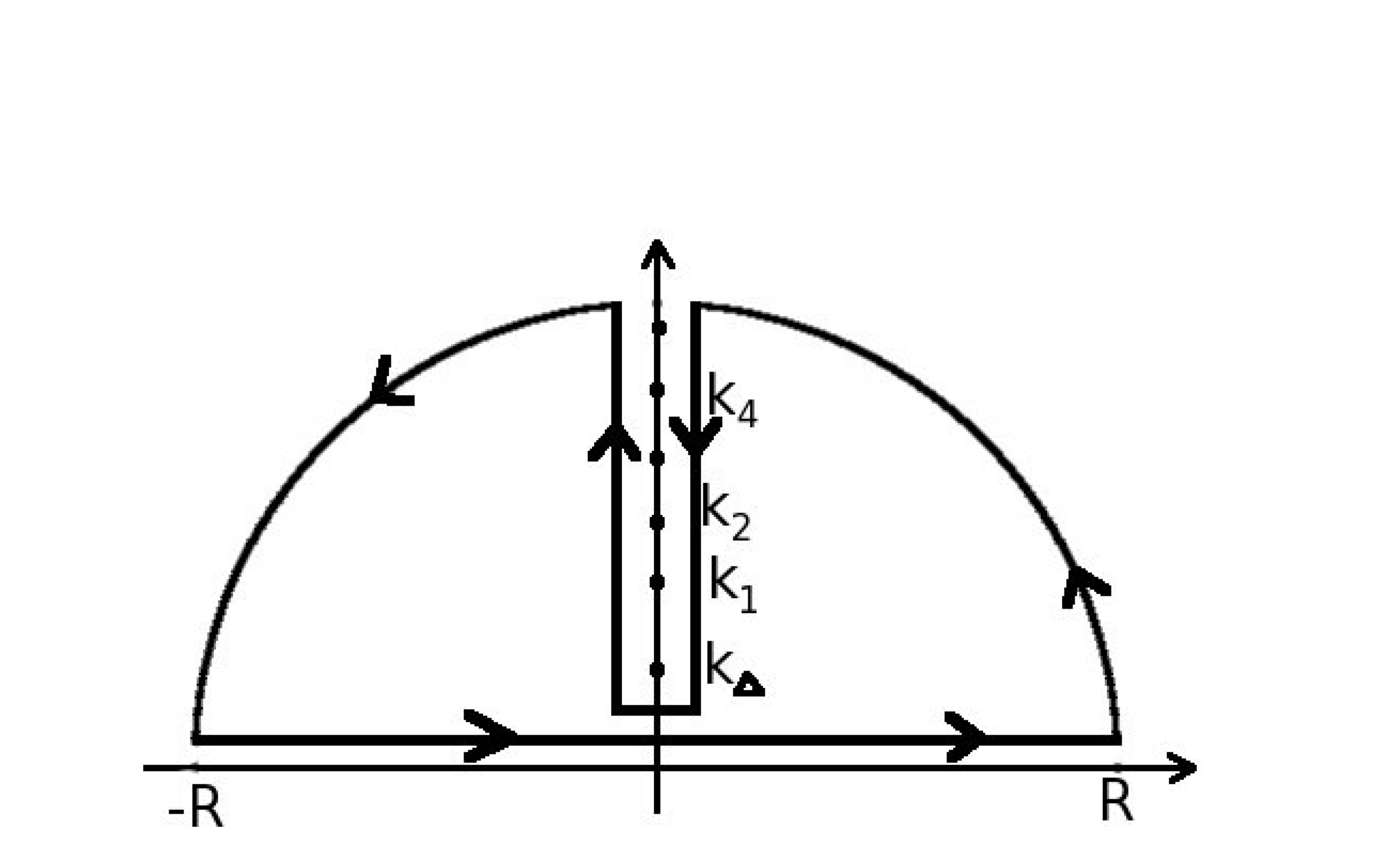}
\caption{\label{contour}
The contour of integration for the functions~(\ref{ep})-(\ref{lm}).}
\end{figure}
We would like to derive the asymptotic behavior of the functions $c_\beta(t)$ and $s_\beta(t)$ determined by Eqs.~(\ref{cbeta}) and (\ref{sbeta}).

To do this, let us first consider the functions:
\bea \label{ep}e^+_\beta(t) = \int_{-\infty}^{\infty} dk \frac{\exp(i k t)}{(k/\Delta)^2+1} \frac{1}{1+\exp(\beta k)},\\ 
 e^-_\beta(t) = \int_{-\infty}^{\infty} dk \frac{\exp(i k t)}{(k/\Delta)^2+1} \frac{1}{1+\exp(-\beta k)}.\eea 
They are defined so that $s=(e^+ - e^-)/2i$.
We define also the functions:
\bea l^+_\beta(t) = \int_{-\infty}^{\infty} dk \frac{k \exp(i k t)}{(k/\Delta)^2+1} \frac{1}{1+\exp(\beta k)}, \\
l^-_\beta(t) = \int_{-\infty}^{\infty} dk \frac{k \exp(i k t)}{(k/\Delta)^2+1} \frac{1}{1+\exp(-\beta k)}.\label{lm}\eea 
Their combination leads to $c=(l^+ - l^-)/2$.

The function under the integral, $f(k)=\frac{1}{1+\exp(\beta k)}\frac{\exp(i k t)}{(k/\Delta)^2+1} $
or $f(k)=\frac{1}{1+\exp(\beta k)}\frac{k\exp(i k t)}{(k/\Delta)^2+1}$,  has poles on the imaginary axis: $k_{\Delta}=\pm i\Delta$, $k_n=i\pi(2n+1)/\beta$, where $n$ is an integer number. 
Let us consider the parameter $t>0$.
We consider the contour $\gamma$ in the upper half-plane, Fig.~\ref{contour}, which goes along a quarter of the circle starting from $|k|=R$, $R\longrightarrow\infty$, then along the imaginary axis down along $Re(k)=0+\delta$, then goes around the pole which is closest to the real axis and then goes up along the imaginary axis at $Re(k)=0-\delta$, continues as a quarter  circle, and then goes to $0+i\delta$ along the real axis. The function $f$ decays uniformly on the quarter-circles at infinity, hence the integrals along the quarter-circles for $R\longrightarrow\infty$ tend to $0$. Therefore, the integral along the real axis is equal to the sum of the residues of the poles:
\be  \label{residues}\int_{-\infty}^{\infty}f(k)dk=\oint_{\gamma} f(k) d \gamma = 2\pi i \sum_{n}\left(\text{Res}_{k_n}f(k)+\text{Res}_{k_{\Delta}}f(k)\right)\ee 

Calculating the values of the residues for $k=i\Delta$ gives (first the value for the function $e$ is given, then for $l$):
\bea r^+_{\Delta| e,l} = &2\pi i& \frac{\Delta\exp(-t\Delta)}{2i}\frac{1}{1+\exp(i \beta\Delta )}, \\ 
&2\pi i& \frac{i\Delta^2\exp(-t\Delta)}{2i}\frac{1}{1+\exp(i \beta \Delta)}\eea
\bea r^-_{\Delta|e,l} = &2\pi i& \frac{\Delta\exp(-t\Delta)}{2i}\frac{1}{1+\exp(-i \beta \Delta)},\\
&2\pi i& \frac{i\Delta^2\exp(-\Delta t)}{2i}\frac{1}{1+\exp(-i \beta \Delta)}
\eea
which adds up to
\bea \frac{1}{\pi\Delta}r_{\Delta,s} = \frac{1}{\pi\Delta}\frac{r^+_{e} - r^-_{e}}{2i} = - \frac{\exp(-t\Delta)}{2} \tan\frac{\beta\Delta}{2},\label{res1}\\
 \frac{1}{\pi\Delta^2}r_{\Delta,c} = \frac{1}{\pi\Delta^2} \frac{r^+_{l}- r^-_{l}}{2} =  \frac{\exp(-t\Delta)}{2} \tan\frac{\beta\Delta}{2} \label{res2}\eea
Notice that the residues for $c$ and $s$ are connected by 
$r_{-1,c}=\frac{d r_{-1,s}}{dt}$ (the same relation holds for the functions itself, $c\Delta=\frac{ds}{dt}$).

The calculation of the residues at $k_n=i\pi(2n+1)/\beta\equiv i\kappa_n$ gives:
\be r^+_{n|e,l} = - \frac{2\pi i}{\beta} \frac{\exp(-t \kappa_n)}{-(\kappa_n/\Delta)^2+1},
 \frac{2\pi}{\beta} \frac{\kappa_n\exp(-t \kappa_n)}{-(\kappa_n/\Delta)^2+1}\ee
\be r^-_{n|e,l} =  \frac{2\pi i}{\beta} \frac{\exp(-t \kappa_n)}{-(\kappa_n/\Delta)^2+1},
-\frac{2\pi }{\beta} \frac{\kappa_n\exp(-t \kappa_n)}{-(\kappa_n/\Delta)^2+1}\ee
and summing them up gives:
\bea r_{n,s}=\frac{r^+_{n|e} - r^-_{n|e}}{2i}= -2 \frac{\pi}{\beta} \frac{\exp(-t \kappa_n)}{-(\kappa_n/\Delta)^2+1}\label{res3} \\
r_{n,c}=\frac{r^+_{n|l} - r^-_{n|l}}{2}= 2 \frac{\pi}{\beta} \frac{\kappa_n\exp(-t \kappa_n)}{-(\kappa_n/\Delta)^2+1}\label{res4}
\eea

The general expression~(\ref{residues}) and the values of the residues (multiplied by $2\pi i$) (\ref{res1},\ref{res2},\ref{res3},\ref{res4}) lead to precise expressions for $c_{\beta}(t)$ and $s_{\beta}(t)$. 
Let us first look at the asymptotic expression of these functions for high temperature $\beta\Delta\ll 1$. In this case the poles of $1/(1+\exp(\beta k))$ lie much further from the real axis compared to the pole at $i\Delta$. The leading term of the sum corresponds to the pole at $i\Delta$, in the subleading terms we expand the fraction $\frac{1}{-(\kappa_n/\Delta)^2+1}$. For the function $s_{\beta}(t)$ we get:
\begin{widetext}
\be s_{\beta\ll1} \approx \frac{1}{\pi\Delta} \left[\frac{\pi \exp(-t\Delta)}{2}\tan\frac{\beta}{2} 
+ \frac{2\beta\Delta^2}{\pi}\sum_{n=0,\ldots}\frac{\exp(-t \kappa_n)}{(2n+1)^2}+ \frac{2\beta^3\Delta^4}{\pi^3}\sum_{n=0,\ldots}\frac{\exp(-t \kappa_n)}{(2n+1)^4} + \Delta O\left((\beta\Delta)^5\right) \right].\ee
The summation for the first subleading term with $\exp\left(-\frac{t\pi}{\beta}\right)$  is given by
\be \frac{2\beta \Delta^2}{\pi} \exp\left(-\frac{t\pi}{\beta}\right) \left(1+\frac{\beta^2\Delta^2}{\pi^2} 
+\frac{\beta^4\Delta^4}{\pi^4} + \ldots\right)= \frac{2\beta\Delta^2}{\pi} \exp\left(-\frac{t\pi}{\beta}\right) \frac{1}{1-\beta^2\Delta^2/\pi^2}\ee
and for the second subleading term is
\be \frac{2\beta\Delta^2}{9\pi} \exp\left(-\frac{3t\pi}{\beta}\right) \left(1+\frac{\beta^2\Delta^2}{9\pi^2} 
+\frac{\beta^4\Delta^4}{9^2\pi^4} + \ldots\right)= \frac{2\beta\Delta^2}{9\pi} \exp\left(-\frac{3t\pi}{\beta}\right) \frac{1}{1-\beta^2\Delta^2/9\pi^2}\ee
etc. Therefore the corrections to the leading term $\frac{\tan(\beta/2)\exp(-t\Delta)}{2\Delta}$ are of the form  $$\sum_{n=0,\ldots} \frac{2\beta}{(2n+1)^2\pi}  \exp(-(2n+1)t\pi/\beta)\frac{1}{1-\beta^2\Delta^2/(2n+1)^2\pi^2}.$$
\end{widetext}
The same expressions divided by the factor $\beta/\pi$ are valid for the function $c_\beta(t)$ . Let us note that this result corresponds with the relation $c\Delta=\frac{ds}{dt}$.

The asymptotic expansions with leading and one subleading term for the functions $c_{\beta}(t)$ and $s_\beta(t)$ are 
\be s_{\beta}(t) \approx - \frac{\exp(-t\Delta)}{2} \tan\frac{\beta\Delta}{2} + \frac{2\beta\Delta}{\pi^2} \exp\left(-\frac{t\pi}{\beta}\right) \frac{1}{1-\beta^2 \Delta^2/\pi^2} \ee
\be c_{\beta}(t) \approx  \frac{\exp(-t\Delta)}{2} \tan\frac{\beta\Delta}{2} - \frac{2}{\pi} \exp\left(-\frac{t\pi}{\beta}\right) \frac{1}{1-\beta^2 \Delta^2/\pi^2}. \ee

Combination of the asymptotic expressions for $c_{\beta}(t)$ and $s_{\beta}(t)$ leads us to the asymptotic expression for the anticommutator $S_{+}(x,t)$,~Eq.(\ref{cplus}).

\section{Expansion for T=0}
For the case $T=0$ we analyze the same expressions (\ref{res1},\ref{res2},\ref{res3},\ref{res4}), performing the summation in~(\ref{residues}) in the limit $\beta\rightarrow\infty$. This leads us to the asymptotic expressions for $s(t)$ and $c(t)$:
\bea s(t) = \frac{1}{t \pi^2 \Delta}\left(1+\frac{2}{t^2\Delta^2}+O\left(\frac{1}{t^4\Delta^4}\right)\right),\\
c(t) = \frac{1}{t^2\pi^2 \Delta^2}\left(1+\frac{6}{t^2\Delta^2}+O\left(\frac{1}{t^4\Delta^4}\right)\right).
\eea
The equal-time anticommutator outside the light cone at zero temperature~(\ref{T0}) follows directly from these expressions.

\end{document}